# Martensite-like transition and spin-glass behavior in nanocrystalline Pr$_{0.5}$Ca$_{0.5}$MnO$_3$

S. Narayana Jammalamadaka[1*], S. S. Rao[2,3*], J. Vanacken[1], A. Stesmans[2], S. V. Bhat[3] and V. V. Moshchalkov[1]

[1]*INPAC – Institute for Nanoscale Physics and Chemistry, Pulsed Fields Group, K.U. Leuven, Celestijnenlaan 200D, 3001 Leuven, Belgium*

[2]*INPAC – Institute for Nanoscale Physics and Chemistry, Semiconductor Physics Laboratory, K.U. Leuven, Celestijnenlaan 200D, 3001 Leuven, Belgium*

[3]*Department of Physics, Indian Institute of Science, Bangalore – 560012, India*

*\* These authors contributed equally to this work*

**Abstract**

We report on isothermal pulsed (20 ms) field magnetization, temperature dependent AC – susceptibility, and the static low magnetic field measurements carried out on 10 nm sized Pr$_{0.5}$Ca$_{0.5}$MnO$_3$ nanoparticles (PCMO10). The saturation field for the magnetization of PCMO10 (~ 250 kOe) is found to be reduced in comparison with that of bulk PCMO (~300 kOe). With increasing temperature, the critical magnetic field required to 'melt' the residual charge-ordered phase decays exponentially while the field transition range broadens, which is indicative of a Martensite-like transition. The AC - susceptibility data indicate the presence of a frequency-dependent freezing temperature, satisfying the conventional Vogel-Fulcher and power laws, pointing to the existence of a spin-glass-like disordered magnetic phase. The present results lead to a better understanding of manganite physics and might prove helpful for practical applications.

# I.   INTRODUCTION

Perovskite manganites having the general formula, $RE_{1-x}AE_xMnO_3$ (where RE is a rare-earth ion and AE an alkaline-earth ion) have attracted considerable scientific and technological interest because of their exotic electronic and magnetic properties[1-3], such as half-metallic nature, colossal magnetoresistance (CMR) and occurance of charge-ordered (CO) phase, to name a few. Most often, the CO phase is associated with an anti-ferromagnetic insulating (AFI) phase, characterized by a long range ordering of the $Mn^{3+}$ and $Mn^{4+}$ ions, due to a complicated competition among Coulomb interactions, exchange interactions, and electron–lattice coupling. The AFI CO phase in bulk (micron size) manganites can be 'melted' (collapsed) into ferromagnetic (FM) metallic (FMM) state by the application of external magnetic fields, electric fields and impurity ion doping, giving rise to CMR[1-3]. To induce CMR in bulk robust charge ordered manganites such as PCMO (x = 0.5), apparently very high magnetic fields (> 270 kOe) are necessary"[4] –a severe constraint to realize CMR effects in practical devices. Hence, a search for new materials is needed.

Intense research efforts have been made [5-10] to investigate the physics of nanoscale CO manganites to explore possible applications. Among the notable findings, weakening of the CO phase and the appearance of a FM phase were reported on several manganites of various CO strengths, though the origin of such particular phenomenon is still under intense debate[5-10]. To quote from archival knowledge[5-10] in explaining such an effect, several groups argued that surface disorder and structural changes upon size reduction could be the main origin. To provide further insights in explaining such puzzling experimental observations, Dong[11] *et al.* have modeled the phenomenon theoretically based on a two-orbital double exchange model near half (0.5) doping using Monte Carlo techniques, concluding that an 'unexpected surface phase separation' could be responsible for the appearance of the weak FM. Later on, it was also shown[12] that the field required to 'melt' the CO phase in nanosized manganites has been reduced drastically from the bulk value ~ 300 kOe to just ~ 50-60 kOe , thereby paving the way to realize prototype CMR-based devices at accessible magnetic fields. A comparative investigation has been conducted on nanowires and nanoparticles of $Ca_{0.82}La_{0.18}MnO_3$, which showed that the nanoparticles of this manganite evidently have different magnetic properties in comparision with the bulk[13].

After the initial report[5] a plethora[12, 14, 15] of experimental observations was reported by several research groups exclusively on nanoscale $Pr_{0.5}Ca_{0.5}MnO_3$ (PCMO), but it's precise magnetic ground state is at present far from being clearly understood. Also the crucial role of size induced disorder needs further exploration. Though a weak FM phase is introduced upon size reduction in such materials[5-10], the 'global FM' state is not achieved, yet an important criterion for spin based applications. Recently, there was a suggestion[15] stating that it is possible to induce a global FM phase by external magnetic fields. However, in trying to realize this, initial[12] isothermal static high magnetic field (H) (up to 14 kOe) – magnetization (M) measurements on PCMO10 failed to achieve the classical saturation magnetization ($M_S$ ~ 3.5 $\mu_B$/f.u.), indicative of global FM.

In the present study, we extended the initially investigated range of magnetic field up to 300 kOe utilizing a nondestructive long-pulse (20 ms) magnet, thereby realizing the much sought saturated magnetization. This occurs at the magnetic fields as high as 250 kOe which is considerably lower than that (~300 kOe) of bulk PCMO. We note that while static magnetic measurements have been used in the past to study the magnetism of nanoparticles of charge ordered manganites, the pulsed high magnetic field studies have not been used to achieve the global FM. In this investigation we specifically address the following issues: (a) realization of global ferromagnetism, (b) the nature of the ground state magnetic phase, in a typical charge ordered manganite such as PCMO upon size reduction.

Our high magnetic pulsed field magnetization and AC - susceptibility measurements reveal that the critical field to 'melt' the (partially) CO phase decreases exponentially with temperature – indicative of Martensite-like transition due to strain developed at the interface between two dissimilar FM-M and AFI-CO phases. The low temperature ground state is inferred to be a spin-glass (SG) like state which might have originated from the competition between the above two distinct magnetic and electronic phases.

## II. **EXPERIMENT**

The sample studied is nanocrystalline $Pr_{0.5}Ca_{0.5}MnO_3$, with an average particle diameter of 10 nm, prepared by the sol-gel method[12] – henceforth refered to as PCMO10. Structural characterization was carried out using the powder x-ray diffraction (XRD) method, by making use of the Rietveld refinement. Transmission electron microscopy (TEM) was used to measure the particle size and material crystallinity.

Preliminary magnetization data have been published elsewhere[12]. The high field magnetization (M-H) measurements were performed at various temperatures on tightly packed PCMO10 particles using the pulsed magnetic field facility at the University of Leuven[16]. We applied pulsed magnetic fields up to 300 kOe with pulse duration of about 20 ms by discharging a capacitor bank through a specially designed magnet coil[17]. For magnetization measurements we used the induction method by employing pickup coils, where the voltage induced in the pickup coils was integrated numerically to obtain the magnetization. For any given measuring temperature, the M-H curve was recorded after first warming up the sample to room temperature (RT) to eliminate remnant effects from previous magnetic observations. The AC-susceptibility measurements were performed using the Physical Property Measurement System (PPMS) in the range T = 2 – 270 K under an AC-magnetic field ($H_{ac}$) of 1 Oe at six difference frequencies in range 13 -1333 Hz. The data were collected during the cooling down run.

The remainder of this paper is organized as follows: In section III(A), we present and discuss the data related to XRD and TEM. The results obtained from pulse field magnetic measurements are presented and discussed in Section III (B); Section III (C) describes the data pertaining to AC susceptibility observations, followed by conclusions in section IV.

## III.    EXPERIMENTAL RESULTS AND DISCUSSION

### A. **XRD and TEM RESULTS**

XRD (indexed) pattern of PCMO10 with Rietveld refinement is shown in Fig. 1a. We could fit the diffraction pattern with a *Pnma* space group symmetry assuming that the Pr, Ca, Mn, O(1) and O(2) atoms are positioned at 4c, 4c,  4b, 4c and 8d sites, respectively, from where it is inferred that the nanoparticles of PCMO10 take the orthorhombic crystal structure (cf. Fig 1.a) with  lattice parameters  a = 5.4734 Å, b = 7.5634 Å, and c = 5.4123 Å and volume V = 224.06 $Å^3$ respectively. From the X-ray data, there is no detectable secondary phase such as $Mn_3O_4$ present. The extracted parameters from Rietveld refinement on PCMO10 as well as the results obtained for bulk $Pr_{0.5}Ca_{0.5}MnO_3$ from neutron diffraction study are comparable with each other[16]. The crystallite size is calculated  using Scherror's formula d = 0.9λ/Bcosθ, where d is the average particle size (nm), λ is the wavelength of the X-rays (1.54 Å), and B is the fullwidth at half maximum of the powder X-ray diffraction peaks;  This gives a particle size of 10 nm. From transmission electron microscopy (TEM) image (cf. Fig. 1b), nearly spherical particles of an average particle size of ~ 10 ±1 nm can be seen, consistent with the particle size obtained from the Scherrer formula.

### B. PULSE FIELD MAGNETIC PROPERTIES

In its bulk form, the narrow bandwidth PCMO exhibits[18] charge ordering (CO) and a charge exchange (CE) type AFM-I phase at $T_{CO}$ = 245 K and $T_N$ = 175 K, respectively. Here, we may add that previous work[12] had found PCMO10 to show interesting phenomena such as 'suppression' of the CO phase, leading to the appearance of a weak FM phase which coexists with a residual CO-AFM phase.

Figure 2 presents the variation of the magnetization as a function of magnetic field applied with a sweep rate of 16 megaOe/s at 8 K. With the application of H, M increases moderately and there is a broad transition observed at around 70 kOe. Under higher magnetic fields, M saturates at ~ 250 kOe where $M_S$ ≈ 3.9 $\mu_B$/f.u. To be noted here is that the field H ≈ 250 kOe required to saturate the magnetization in PCMO10 is considerably reduced compared with that (~300 kOe) of its bulk counterpart[4], consistent with the recent theoretical predictions[11] [see, Fig 3 of Appl. Phys. Lett., 90, 082508 (2007)]. Such a reduction in the saturation field might be partially due to size induced disorder, resulting in the reduction of the free energy difference between the predominant FMM and residual AFM COI phases.

To track the temperature dependence of this field induced transition, isothermal M-H measurements were performed at reduced field sweep rate in the field range of 0→170 kOe (of 20 ms pulse duration) at various temperatures so as to clearly observe the transition. Figures 3 (a-f) present the variation of M as a function of H sweep (0 →170 kOe→0) measured at six temperatures. From these plots, several interesting features can be noticed. While H is ramping up (0 →170 kOe), M increases moderately up to a certain H value ≈ 50 kOe, to increase from there on at a higher rate. It signals field induced transition from the residual AFI-CO to the FMM state[12]. The reverse cycle (170 kOe→0) trace does not overlap with the warm up curve indicating a substantial magnetic hysteresis[12], the effect being more pronounced at lower T. The observed M-H loop area gradually closes as the sample is warmed up to 70 K. Similar features were earlier reported[12] by conventional DC SQUID measurements performed on PCMO20, PCMO40, on $Nd_{0.5}Ca_{0.5}MnO_3$ nanoparticles[19] as well as on Al doped bulk PCMO[20]. Also, much sharper stair-case like metamagnetic transitions were seen in several other Pr and Nd based bulk manganites[21,22]. The possible reason[21] for the current broadened transition could be that the spins are strongly coupled to the lattice, hindering the kinetics of AFM to FM phase transitions. The sharpness of this transition also depends upon the field sweep rate and disorder present in the compound, and is the subject of future work.

The transition field ($H_j$), at which M rises faster, is determined accurately by computer fitting of the first derivative of the pristine $M - H$ curve (0→170 kOe) to a Gaussian peak at each temperature. The results are plotted in Fig. 4, from where it is clearly seen that $H_j$ is a rather sensitive function of temperature, decaying exponentially as indicated by the dashed curve. The transition is broadened beyond detection for T above 70 K. A similar $H_j - T$ phase diagram, where $H_j$ is a strong function of T (exhibiting an exponential decay), was reported[22] earlier in Cr -doped $Nd_{0.5}Ca_{0.5}MnO_3$, attributed to Martensite-like transition. Likewise, the current $H_j - T$ data for PCMO10 also signal the presence of features typical for a Martensite-like transition. However, within the Martensite-like scenario, it can also be due to the collapse of the balance between the magnetic energy and the elastic energy under an ultrafast field that results in the suppression of the magnetization steps. The present features could also correspond to the pseudo CE-type AFM to FM transition[23]. In the disordered (due to size reduction) manganites, the slight mismatch in the unit cell parameters of competing FMM and AFI-CO phases generates a notable strain at the interface regions, creating a situation similar as for the strain accommodation observed in classical Martensite phase transition[24], which may explain the current behaviour.

## C. AC SUSCEPTIBILITY PROPERTIES

Now we focus on the interpretation of the results obtained from AC-susceptibility measurements performed to probe the dynamics of the spin system. The magnetization can be expressed in terms of magnetizing field as

$$M = M_0 + \chi_1 H + \chi_2 H^2 + \chi_3 H^3 + ..........,$$  (1)

where $M_0$ is the spontaneous magnetization, $\chi_1$ is the linear term and $\chi_2$, $\chi_3$, are the 2nd and 3rd harmonics. Figure 5 (a) presents the linear response of the susceptibility $\chi_1$ (T) in the range T = 2-270 K measured at six fixed frequencies on PCMO10. The suppression of the CO phase is evidenced from the absence of a peak in the $\chi_1$ (T) plot at around $T_{CO} \sim 245$ K, and the occurance of a steep rise in $\chi_1$ (T) for T decreasing below ~150 K, which indicates the appearance of a FM phase, well in agreement with existing reports[5-15]. At high temperatures (>200 K), $\chi_1$ (T) is found to follow a Curie-Weiss (C-W) behavior with $T_{C-W} = 75 \pm 5$ K. This value, obtained by extrapolating the linear part to $\chi^{-1} = 0$, is positive ($T_{C-W} > 0$) for all frequencies. Corresponding effective magnetic moment is 4.41 $\mu_B$/f.u. This value is found to be larger than spin only effective magnetic moment (3.5 $\mu_B$) for PCMO (x = 0.5). The additional magnetic moment might have originated from the paramagnetic contribution of $Pr^{3+}$ or it may reflect the evidence of magnetic polarons in the paramagnetic phase. As the sample is cooled further down, a peak in $\chi_1$ (T) appears at T ≈ 40 K for all frequencies.

Upon closer inspection, as shown in the inset of Fig. 5a, the peak value is found to shift to higher temperatures with increasing frequency −a bench mark signature of SG behaviour[25]. As the SG and the superparamagnetic state (SPM) share similar experimental features, it then becomes quite complex to pin-point the exact magnetic nature of the system. To discriminate between these possible SG or SPM phases, as mentioned earlier, we have also performed temperature dependent measurements of the nonlinear AC susceptibility $\chi_2$ (T) and $\chi_3$ (T) (not shown). Usually, $\chi_2$ appears due to the presence of a symmetry breaking field which originates either from a spontaneous magnetization or a superimposed dc magnetic field. In our nonlinear AC susceptibility results, no peak is observed in $\chi_2$ (T) nor in $\chi_3$ (T), which would indicate the absence of spontaneous magnetization −typical for a SG-like system. This analysis would thus indicate that PCMO10 shows SG−like cooperative freezing, arising from size induced predominant FMM and residual AFM COI phases.

In order to confirm the SG-like behavior more quantitatively, and to gain further insight and confidence, we here calculate a dimensionless parameter, g, serving as a decisive factor in favor of either SG or SPM, given as[25]

$$g = \frac{\Delta T_{SG}}{T_{SG}\Delta\log(f)} \quad,$$
(2)

In this equation, $\Delta$ refers to the difference between the measurements at different frequencies. For a SG system, g is of the order of 0.01 while for SPM, g > 0.1. For PCMO10, we find g ≈ 0.0125, which falls in the range of a SG system, thus giving further support to the SG nature of the magnetic state in PCMO10.

The origin of the temperature and frequency dependences of freezing temperature ($T_f$) in a genuine SG has been the subject of intense debate during the last decades[26-27]. However, the dynamic scaling theory[26] is generally admitted to be the most relevant to account for the SG transition[26]. This theory predicts a power law of the form

$$\tau = \tau_0 [\frac{(T_f - T_{SG})}{T_{SG}}]^{-zv} \quad , \qquad (3)$$

where, $\tau$ is the spin relaxation time, $\tau_0$ is the shortest spin relaxation time available in the system, $T_{SG}$ is the underlying spin-glass transition temperature determined by the interactions in the system, $z$ is the dynamic exponent, and $v$ is the critical exponent of the correlation length. To obtain the dynamical critical exponents ($z$ and $v$) and the spin relaxation time ($\tau_0$), first, the value of the spin-glass transition temperature, $T_{SG}$, was adjusted to get the best linearity in the log(f) versus log[(T-$T_{SG}$)/$T_{SG}$] plot (cf. Fig. 5(b)). The values of best fitting parameters thus obtained for PCMO10 are $T_{SG}$ = (37.0 ± 0.2) K, $\tau_0$ = (1.1 ± 0.3) x10$^{-10}$ sec and $zv$ = 9.0 ± 0.4. The obtained values of $\tau_0$ and $zv$ fall within the range of values typical for spin glasses[26] i.e, 10$^{-10}$ – 10$^{-12}$ s and 5-10, for $\tau_0$ and $zv$ ,respectively. As expected for dynamic scaling theory, the value of $T_{SG}$ is close to the location of the maximum in the temperature-dependent zero field cooled magnetization $M_{ZFC}$ (T) curve (cf. Fig.1a. of Ref 12).

In a next step, we seek to corroborate the observation of the SG-like phase in PCMO10 by existing models. For a system possessing interparticle interactions such as dipolar and exchange interactions, SG dynamics can be governed by the Vogel-Fulcher law[27]

$$T_f = T_0 + \frac{E_a}{k_B [\ln(\frac{f_0}{f})]^{-1}} \quad ., \qquad (4)$$

where $T_f$ is the spin-freezing temperature, $T_0$ the Vogel-Fulcher temperature, $E_a$ the activation energy for relaxation, $k_B$ is the Boltzmann constant, $f_0$ is a characteristic frequency and f the applied frequency. The Vogel-Fulcher law is found to be applicable for PCMO10 as well, as illustrated by the "straight line" behaviour in Fig. 5(c), where the best fitting gives $E_a/k_B$ =175 K, $T_0$ = 32 K, and $f_0$ = $10^{10}$ Hz.

The observation of a SG behavior can be understood assuming that the relaxation of the superexchange interaction at the surface of nanoparticles allows the formation of a FM or SG shell, which can lead to the formation of AFM/FM or FM/SG interfaces[11, 28]. In addition, the nanoparticles have a high surface - to - volume ratio which can result in an uncompensated spin and, as a result, suppression of the long-range AFM order observed in the bulk.

## IV. **CONCLUSIONS**

Upon the application of high pulsed magnetic fields on PCMO10 using a set up PFM, we find that the long sought 'global ferromagnetism' could be achieved at magnetic fields of $\approx$ 250 kOe, considerably lower than ($\sim$ 300 kOe) characteristic for bulk PCMO. Consequently, we conclude an evidence for the occurrence of a Martensite-like transition as supported by the observation of an exponential decay of the critical field with increasing temperature in PCMO10.

The AC - susceptibility measurements reveal spin-glass-like behavior at low temperatures with $T_{SG}$ = 37 K, satisfying a conventional power law as well as the Vogel- Fulcher law. While the former transition may arise from the natural consequence of strain at the interface of FMM and AFM-CO phases in PCMO10, the latter phase is the result of magnetic frustration.

**ACKNOWLEDGMENTS**

SNJ would like to thank KU Leuven, for research fellowship, SSR acknowledges CSIR, Government of India for fellowship. SVB thanks DST, India for project funding under NSTI. The authors would like to thank Alejandro Silhanek for his help during the measurements. This work is supported by the K.U. Leuven Excellence financing (INPAC), by the Flemish Methusalem financing and by the IAP network of the Belgian Government.

*Corresponding authors

Surya.Jammalamadaka@fys.kuleuven.be  and srinivassingamaneni@boisestate.edu

therein.

**Figure captions**

Fig. 1: (a) Rietveld refined XRD pattern of PCMO10 (b) TEM micrograph of PCMO10

Fig. 2: The variation of magnetization (M) as a function of applied magnetic field measured at T = 8 K on PCMO10, with a pulse duration and pulse sweep rate of 20 ms and 16 megaOe/s, respectively. One can notice that the magnetization tends to saturate at around 250 kOe (indicated by dotted line).

Fig. 3: (Color online) Isothermal M-H curves observed on PCMO10 for H up to 170 kOe at various temperatures. The indication of a field induced magnetic phase transition is observed as a kink at each temperature in the warm up curve. For T > 70 K, the transition width broadens further beyond detection. It is apparent that the critical field ($H_j$) increases with lowering temperature. The direction of the field sweep is indicated by arrows.

Fig.4: Thermal evolution of the critical field ($H_j$) obtained from Gaussian fitting of the first derivative dM/dH of the observed M-H traces, revealing an exponential decay, as indicated by the dashed curve.

Fig. 5 (Color online) (a) Temperature dependence of the real part (in-phase) of the a.c. susceptibility ($\chi_1$) measured at different frequencies under an applied $H_{a.c}$= 1 Oe, exhibiting a peak at around T ~ 40 K. The corresponding inset shows an expanded version of the peak T range at around T ~ 40 K to analyze the frequency dependence of peak shift; this is seen to shift to higher temperature with f (cf. arrow in plot), a fingerprint signature of a SG phase; (b) Plot of f versus [(T-$T_{SG}$)/ $T_{SG}$] in log-log coordinates. The straight line represents the optimized computer fitting result giving $T_{SG}$ = (37 ± 0.2) K, $\tau_0$ = (1.1 ± 0.3) x$10^{-10}$ and $zv$ = 9.0 ± 0.4, indicating that the SG behaviour is governed by a power law. (c) The f dependence of $T_f$ in $T_f$

versus $[\ln(f_0/f)]^{-1}$ coordinates, where the straight line is obtained from computer fitting of the Vogel- Fulcher law, with best fitting values obtained as $E_a$ =175 K, $T_0$ = 32 K and $f_0$ = $10^{10}$ Hz.

**FIGURES**

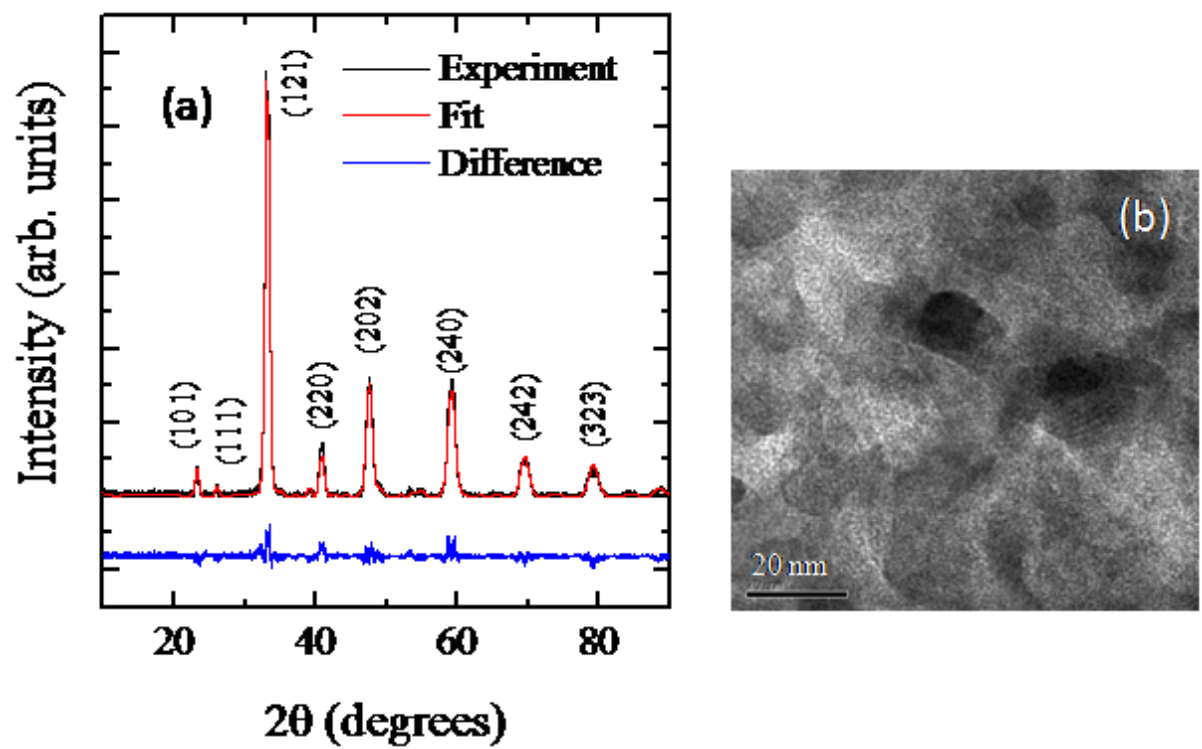

Fig. 1: (Color online) (a) Rietveld refined XRD pattern of PCMO10; (b) TEM micrograph of PCMO10

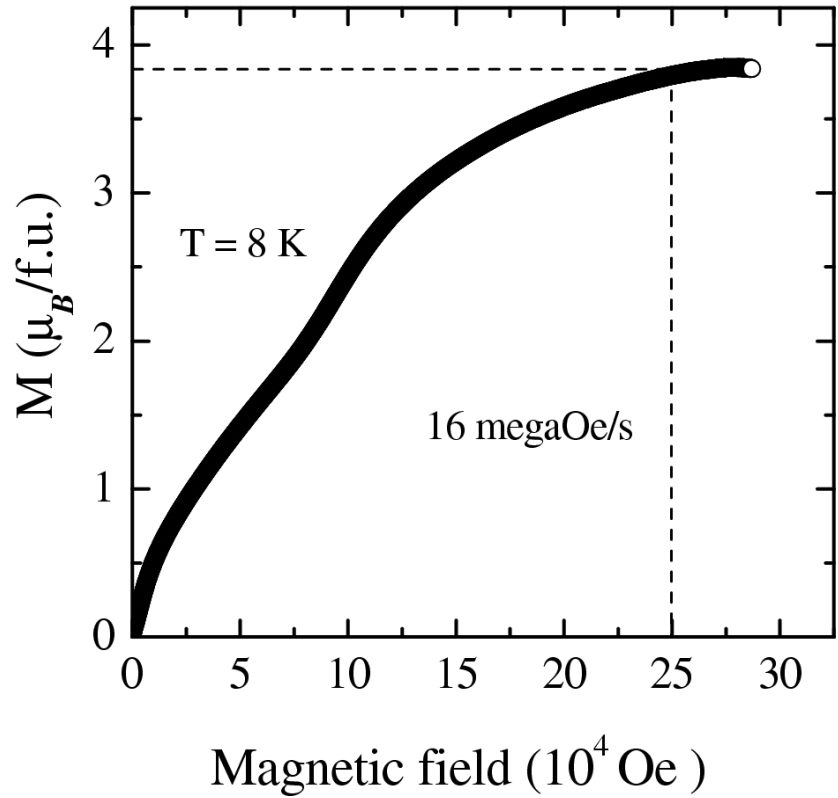

Fig. 2: The variation of magnetization (M) as a function of applied magnetic field measured at T = 8 K on PCMO10, with a pulse duration and pulse sweep rate of 20 ms and 16 megaOe/s, respectively. One can notice that the magnetization tends to saturate at around 250 kOe (indicated by dotted line).

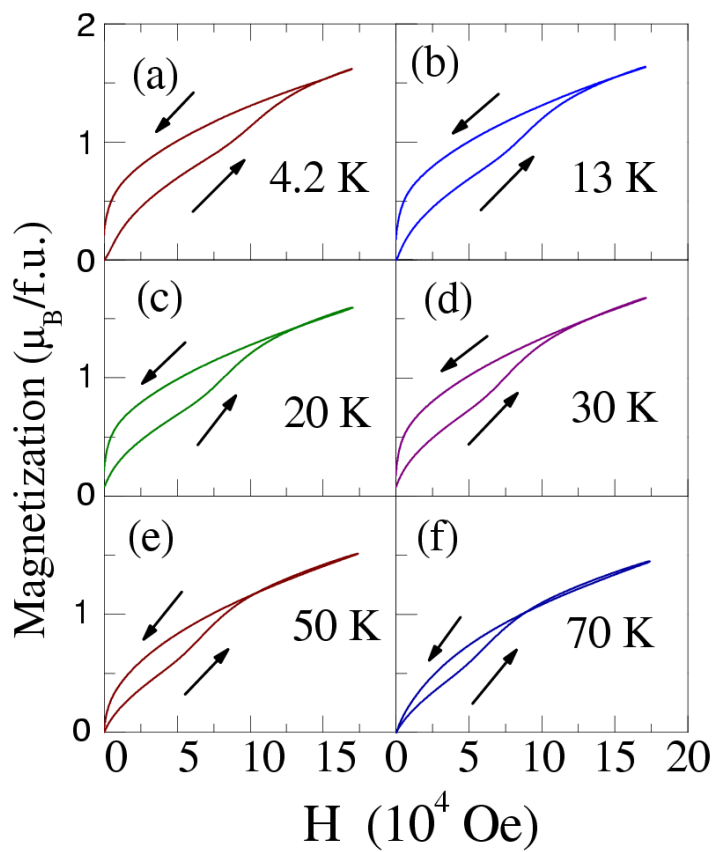

Fig. 3: (Color online) Isothermal M-H curves observed on PCMO10 for H up to 170 kOe at various temperatures. The indication of a field induced magnetic phase transition is observed as a kink at each temperature in the warm up curve. For T > 70 K, the transition width broadens further beyond detection. It is apparent that the critical field (H$_j$) increases with lowering temperature. The direction of the field sweep is indicated by arrows.

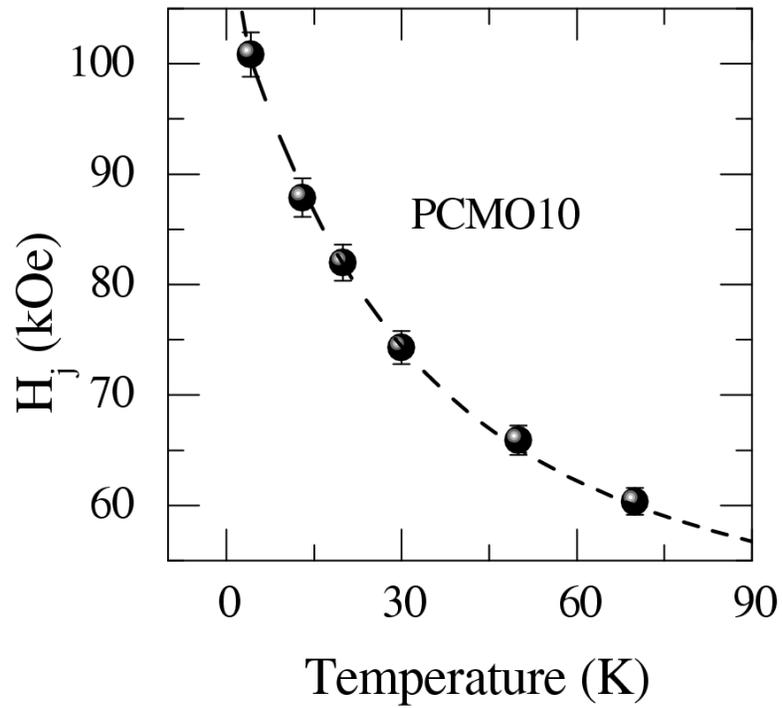

Fig.4: Thermal evolution of the critical field ($H_j$) obtained from Gaussian fitting of the first derivative dM/dH of the observed M-H traces, revealing an exponential decay, as indicated by the dashed curve.

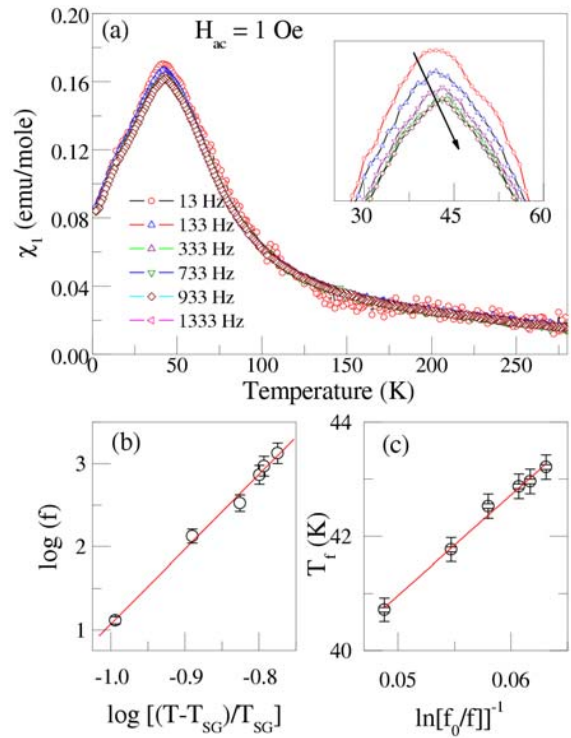

Fig. 5 (Color online) (a) Temperature dependence of the real part (in-phase) of the a.c. susceptibility ($\chi_1$) measured at different frequencies under an applied $H_{a.c}$= 1 Oe, exhibiting a peak at around T ~ 40 K. The corresponding inset shows an expanded version of the peak T range at around T ~ 40 K to analyze the frequency dependence of peak shift; this is seen to shift to higher temperature with f (cf. arrow in plot), a fingerprint signature of a SG phase; (b) Plot of f versus $[(T-T_{SG})/ T_{SG}]$ in log-log coordinates. The straight line represents the optimized computer fitting result giving $T_{SG}$ = (37 ± 0.2) K, $\tau_0$ = (1.1 ± 0.3) x$10^{-10}$ and $zv$ = 9.0 ± 0.4, indicating that the SG behaviour is governed by a power law. (c) The f dependence of $T_f$ in $T_f$ versus $[\ln(f_0/f)]^{-1}$ coordinates, where the straight line is obtained from computer fitting of the Vogel- Fulcher law, with best fitting values obtained as $E_a$ =175 K, $T_0$ = 32 K and $f_0$ = $10^{10}$ Hz.